\documentclass[aps,prl,twocolumn,amsmath,amssymb, superscriptaddress]{revtex4-2}
\usepackage{graphicx}% Include figure files
\graphicspath{{Figures/}}
\usepackage{subfigure}
\usepackage{epsfig}
\usepackage{ulem}
\usepackage{dcolumn}% Align table columns on decimal point
\usepackage{bm}% bold math
\usepackage{hyperref}% add hypertext capabilities
\hypersetup{colorlinks=true, citecolor=blue, urlcolor=blue, linkcolor=blue}
\IfFileExists{newtxtext.sty}
{\usepackage{newtxtext,newtxmath}}
{\IfFileExists{stix.sty}
	{\usepackage{stix}}
	{\IfFileExists{mathptmx.sty}
		{\usepackage{mathptmx}}{} } }

\usepackage{textcomp}
\usepackage{bm}

\IfFileExists{siunitx.sty}{\usepackage{booktabs,siunitx}}{}

\usepackage{color}
\definecolor{LinkColor}{rgb}{0.256,0.439,0.588}
\usepackage{hyperref}
\hypersetup{
	pdfauthor={good guys},
	pdftitle={good title},
	colorlinks=true,
	citecolor=LinkColor,
	linkcolor=LinkColor,
	urlcolor=LinkColor
}

\newcommand{\beq} {\begin{equation}}
\newcommand{\eeq} {\end{equation}}
\newcommand{\bea} {\begin{eqnarray}}
\newcommand{\eea} {\end{eqnarray}}
\newcommand{\be} {\begin{equation}}
\newcommand{\ee} {\end{equation}}

\newcommand{\ket}[1]{\left|#1\right>}
\newcommand{\bra}[1]{\left<#1\right|}
\def\avg#1{\left\langle#1\right\rangle}

\def\Eq#1{Eq.~(\ref{#1})}
\def\Fig#1{Fig.~\ref{#1}}

\begin{document}
\title{Emergence of Competing Orders and Possible Quantum Spin Liquid in  SU(N) Fermions}
\author{Xue-Jia Yu}
\affiliation{International Center for Quantum Materials, School of Physics, Peking University, Beijing 100871, China}
\author{Shao-Hang Shi}
\affiliation{Beijing National Laboratory for Condensed Matter Physics and Institute of Physics,
Chinese Academy of Sciences, Beijing 100190, China}
\affiliation{University of Chinese Academy of Sciences, Beijing 100049, China}
\author{Limei Xu}
\affiliation{International Center for Quantum Materials, School of Physics, Peking University, Beijing 100871, China}
\affiliation{Collaborative Innovation Center of Quantum Matter, Beijing, China}
\affiliation{Interdisciplinary Institute of Light-Element Quantum Materials and Research Center for Light-Element Advanced Materials, Peking University, Beijing, China}
\author{Zi-Xiang Li}
\email{zixiangli@iphy.ac.cn}
\affiliation{Beijing National Laboratory for Condensed Matter Physics and Institute of Physics,
Chinese Academy of Sciences, Beijing 100190, China}
\affiliation{University of Chinese Academy of Sciences, Beijing 100049, China}

\begin{abstract}
{In the past decades, tremendous efforts have been made towards understanding the exotic physics emerging from competition between various ordering tendencies in strongly correlated systems.  Employing state-of-the-art quantum Monte-Carlo simulation, we investigate an interacting SU($N$) fermionic model  with varying interaction strength and value of $N$, and unveil the ground-state phase diagram of the model exhibiting a plethora of 
exotic phases. For small value of $N$, namely $N=2,3$, the ground state is antiferromagnetic (AFM) phase, whereas in the large-$N$ limit, valence bound solid (VBS) order is dominant. For the intermediate value of $N$ such as $N=4$, remarkably, our study reveals distinct VBS orders appear in the weak and strong coupling regimes. More fantastically, the competition between staggered and columnar VBS ordering tendencies gives rise to a Mott insulating phase without spontaneously symmetry breaking (SSB), existing in a large interacting parameter regime, which is consistent with a gapped quantum spin liquid. Our study not only provides a platform to investigate the fundamental physics of quantum many-body systems, but also offers a novel route towards searching for exotic states of matter such as quantum spin liquid in realistic quantum materials.  }
\end{abstract}
\date{\today}
\maketitle

\begin{figure}[t]
\includegraphics[width=0.5\textwidth]{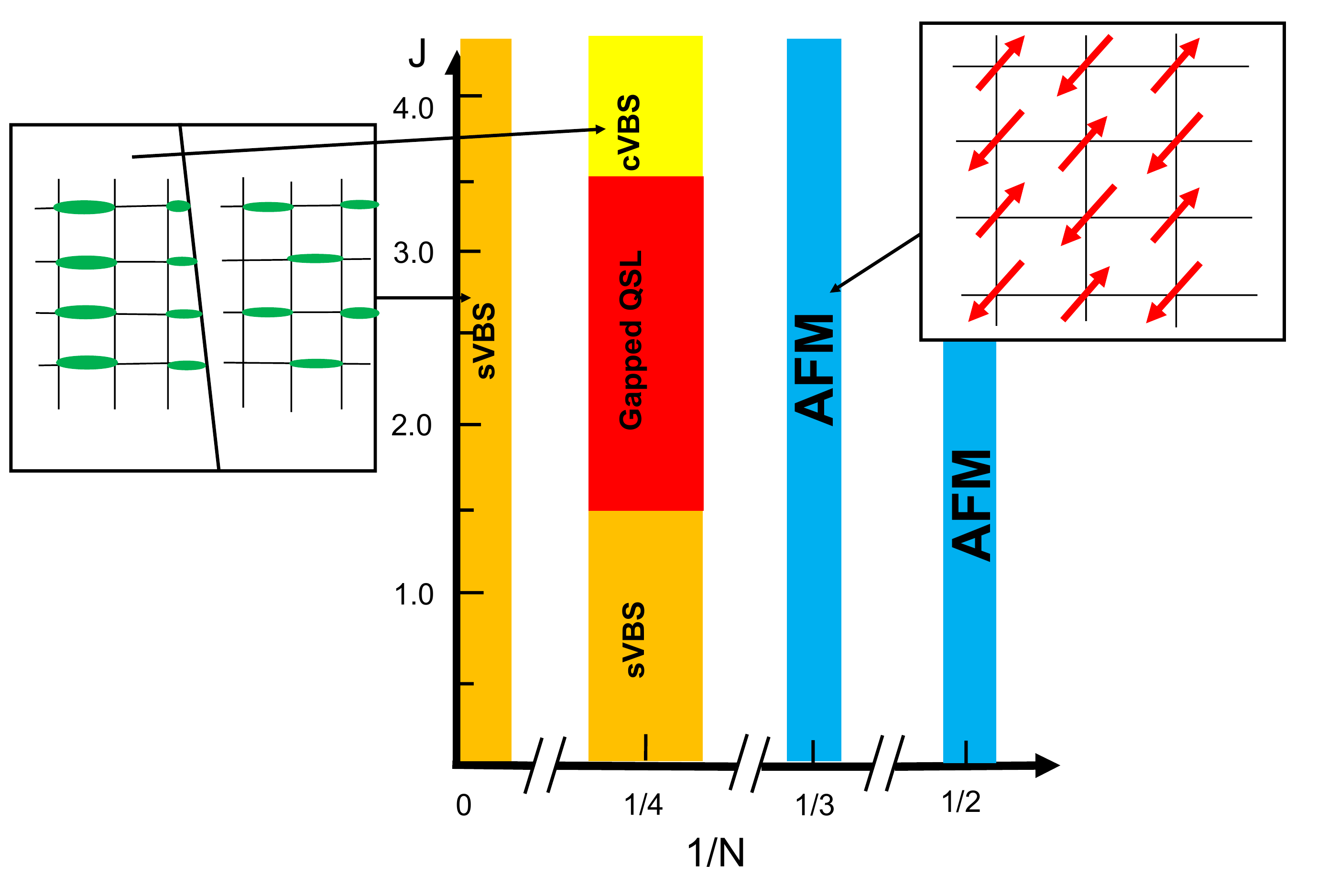}
\caption{Schematic ground-state phase diagram of SU($N$) fermions with SSH interaction on square lattice at half filling: for $N = 2, 3$, the ground state is AFM ordered phase.  In the large $N$ limit, the dominant VBS ordering is sVBS with momentum $(\pi,\pi)$. For the intermediate $N$, such as $N = 4$, the insulating phases in weakly and strongly interacting regime possess VBS orders with distinct patterns of SSB. The sVBS order carrying momentum $(\pi,\pi)$ is dominate in the weakly interacting regime owing to the presence of Fermi-surface nesting, while in strongly coupling regime, cVBS order with momentum $(\pi,0)/(0,\pi)$ appears. Most appealingly, in a large regime between two different VBS phases ($1.5<J<3.5$), the ground state is a Mott insulator without SSB consistent with gapped quantum spin liquid. }
\label{fig1}
\end{figure} 

{{\it Introduction.-}} Understanding competing orders arising from strong electronic interactions plays vital roles in fathoming fundamental theory of quantum many-body physics~\cite{sachdev_2011,fradkin2013field}, as well as many appealing phenomena in quantum materials including high $T_c$ superconductivity~\cite{Lee2006rmp,FradkinReview,SachdevReview}. More intriguingly, the competitions between various ordering tendencies offer a route towards yielding exotic states of matter. A typical example is quantum spin liquid  (QSL)~\cite{Balents2010Review,Lee2007Review,zhou2017rmp,Savary_2016,NormanReview,KivelsonReviewQSL}, a Mott insulating phase featuring deconfined fractionalization, whereas without symmetry spontaneously breaking even at zero temperature. Despite vigorous studies in past decades~\cite{Wen2004PRB,Fisher2005PRB,yan2011spin,Jiang2012PRB,Han2012Nature,Saito2011Nature,meng2010quantum,sorella2012absence,Xiang2017PRL,Yao2007PRL,Yao2011PRL,Jiang2019PRB,Sheng2015PRB,Zhou2008PRL,he2017prx,Zhou2022SB,wang2021prb,yan2021topological,Wan2014PRB,Song2019NC,wang2018prl,aaron2020prx,Li2022PRB}, unambiguous demonstration of realizing quantum spin liquid in realistic microscopic models by unbiased theoretical approach remains elusive.

On a different front, another promising mechanism to trigger exotic state of matter is assigning more degrees of freedom to the microscopic models  and extending symmetries. By extending spin symmetry group SU($2$) in electronic spin system to SU($N$), it is theoretically predicted that quantum fluctuation of spin is enhanced with increasing $N$\cite{read1989some,read1989prl}, rendering the possibility of exotic quantum paramagnetic phases such as valence bond solid (VBS)\cite{lang2013prl,Assaad_SUN2005,Kaul2012PRL,zhou2016prb} and quantum spin liquid\cite{corboz2012prx,koga2018prb,Zhou2022SB}. Fruitful fascinating physics has been revealed in the interacting SU($N$) systems by virtue of the 
cooperative effects of enlarged symmetry and strong correlation ~\cite{Honerkamp2004PRL,chen2021mott,xu2019mott,cai2020symplectic,wang2019prb,zhou2018prb,xu2018prl,Kaul2013Review,Mila2016PRL,zhou2014prb,wang2014prl,cai2013prb,cai2013prl,Mila2014PRL,hung2011prb,marcin2020prr,kim2019prb,Liao2022prb,Liao2022PRB2,zhu2022prb,Alet2009PRB,Troyer2003PRL,Chen2022PRR,assaad2016prx,lukas2022prb,kazuki2207,wang2208,ibara20210pra,hironobu2021prl,romen2020prr,fromholz2020prb,Qi2022PRB}. Benefiting from celebrated progresses in cold-atom experiments, various SU($N$) 
symmetric systems are realized in optical lattice \cite{desalvo2010prl,Gorshkov2010NP,taie2010prl,krauser2012coherent,taie20126,coldatom4,coldatom5}. More recently, enlarged SU($N$) symmetry is proposed to emerge in the twisted bilayer graphene system ~\cite{Cao2018Nature1,Cao2018Nature2,Macdonald2013PNAS} as a low-energy effective description combining electronic spin and valley degrees of freedom\cite{Chubukov2022PRL,xu2018prb,liao2019prl,You2019Npj}.  Hence, designing concrete model with extended symmetry and investigating the effect of strong correlation in a theoretically controlled approach is vastly desired.

In this letter, we construct a SU($N$) interacting fermionic model on square lattice. Remarkably, the model is sign problem free at half filling for any $N$\cite{zixiang2015prb,zxli2016majorana,Wang2015PRL,Xiang2016PRL}, thus amendable to approximation-free Quantum Monte Carlo (QMC) simulation~\cite{assaad2008world,bss1981prd,Li2019Review} at low temperature (or zero temperature) with large system sizes.  We perform large-scale projector QMC simulation and investigate numerically accurate ground-state properties of the model with varying interaction strength and value of $N$. At small value of $N$, the ground state is AFM spin ordered state, like SU(2) Heisenberg model on square lattice.  With increasing value of $N$, quantum fluctuation of spin destroys spin long-range order, rendering VBS ordering breaking translational symmetry, whereas preserving spin rotational symmetry. The most intriguing phenomena occur in the intermediate value of $N$, wherein the insulating phases in weakly coupling regime and strongly coupling regime feature distinct patterns of VBS. The competition between two ordering tendencies gives rise to an intermediate insulating phase without any SSB even at zero temperature. Such symmetric Mott insulator is a possible gapped quantum spin liquid phase featuring topological order~\cite{read_sachdev_1991prl,wen1991prb,chen2010prb,wen1990topological}.

{{\it Model and Method.-}} We consider the following interacting microscopic model of SU($N$) fermions on the square lattice:\cite{li2017fermion}(SSH interaction model):
\begin{equation}
\label{E1}
H=-t\sum_{\langle ij \rangle,\alpha} (c^{\dagger}_{i\alpha}c_{j\alpha}+H.c.)- \frac{J}{2N}\sum_{\langle ij \rangle}(\sum_\alpha c^{\dagger}_{i\alpha}c_{j\alpha}+H.c.)^{2}
\end{equation}
where $\langle ij \rangle$ refers to the bond between nearest-neighbor (NN) sites $i$ and $j$, $c^{\dagger}_{i\alpha}$ creates a fermion on-site $i$ with fermion flavor label $\alpha=1,...N$. Here $t$ is the fermion hopping amplitude and is set $t=1$ as an energy unit. $J$ represents the strength of the SSH interaction, which can be induced by SSH electron-phonon couplings in the fast phonon limit\cite{ssh1979,fradkin1983}. Remarkably, model in \Eq{E1} is free from the notorious sign problem for any $N$ at half filling and even $N$ at generic filling, and thus a promising platform to investigate the exotic physics emerging in the SU($N$) fermions. In this work, we perform un-biased, large-scale projector QMC to study the ground-state properties of the model. The details of projector QMC are introduced in the Supplementary Materials (SM). We focus on the model at half filling with perfectly nested Fermi surface, and explore the insulating phase with competing ordering arising from the interplay between strong correlation and extended symmetry of multi-component fermions. 

Notice that for $N=1$, the interaction in \Eq{E1} is reduced to the nearest-neighbor density interaction, which favors charge-density-wave order on bipartite lattice, as intensively studied previously\cite{Wang2014NPJ,Li2015NPJ,Li2020PRB,Guo2020PRB}. For $N=2$, the model is SSH electron-phonon coupling model at anti-adiabatic limit, the ground-state of which is revealed to be AFM ordered state in recent works~\cite{cai2021prl,Cai2022prb,gotz2022prb,feng2022prb}. In the present work, we focus on understanding the nature of the multi-component fermions with enlarged symmetry, namely $N \geq 3$, in the presence of strong correlation.

{{\it Quantum phase diagram.-}} 
Before presenting the details of QMC results, we summarize our main findings and the ground-state phase diagram of the model in \Eq{E1}. The schematic phase diagram with varying fermion-flavor number $N$ is shown in \Fig{fig1}. For small $N$, more explicitly $N=2,3$, the ground state is an insulating phase with AFM spin long-range order. As discussed in previous literature, with increasing fermion-flavor number $N$, the quantum fluctuation of spin is enhanced, resulting in the emergence of valence bond ordering. The emergence of VBS order for large-$N$ case is revealed by our systematic numerical calculation. In the large-$N$ limit, where the saddle-point approximation can render the accurate results, we perform mean-field calculation and unambiguously show that the ground state is VBS ordered phase with momentum $(\pi,\pi)$, namely a staggered VBS (sVBS) phase. In the case of a large but finite $N$, we perform QMC simulation to decipher the rich ground-state phase diagram. For example, for $N=4$, the AFM spin order is suppressed and VBS order becomes dominant. More strikingly, our state-of-the-art QMC simulation shows that the patterns of VBS orders in the weak and strong coupling regimes are completely different. In the weakly interacting regime, the VBS order in the ground state carries momentum $(\pi,\pi)$ as a consequence of Fermi surface nesting. In the strongly coupling regime, the dominant pattern of VBS ordering is the colomnar VBS (cVBS) with momentum $(\pi,0)/(0,\pi)$. Appealingly, in a large regime with intermediate interacting strength ($1.5<J<3.5$), a phase without SSB arises from the competition between two VBS ordering tendencies. This phase without any SSB in the intermediate strength of $J$ possesses finite single-particle and spin gaps. Therefore, it is a Mott insulating phase, and possibly, a spin liquid phase with exotic topological order.

\begin{figure}[tb]
\includegraphics[width=0.5\textwidth]{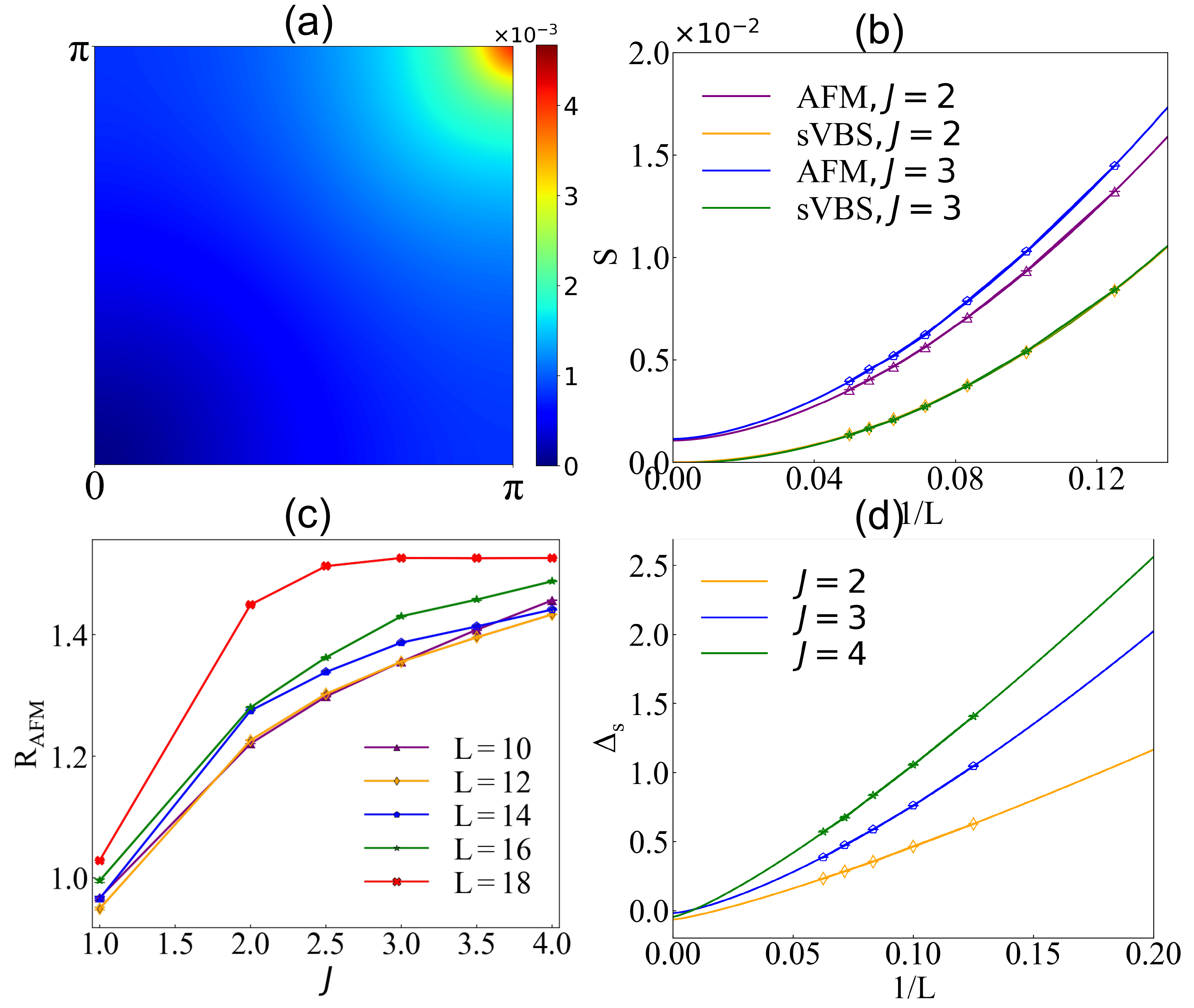}
\caption{Numerical evidence of AFM order in $N=3$: (a) The color plot of momentum distribution of AFM structure factor for $J=2, L=16$, clearly showing that peak occurring at momentum $(\pi,\pi)$. (b) Finite-size scaling of spin and VBS structure factor at peaked momentum $(\pi,\pi)$. The second-order polynomial is used to fit the structure factor as a function of $1/L$, and the intercept of extrapolation indicates the structure factor at thermodynamic limit. (c) RG-invariant AFM correlation-length ratio as a function of interacting strength $J$ for system size $L=10,12,14,16$. Increasing correlation ratio with system size is a strong evidence of long-range order. (d) The extrapolation of spin gap as a function of $1/L$ for $J=2,3,4$. Second-order pronominal is used to fit the structure factors. The gapless spin excitation is consistent with AFM long-range ordering. }
\label{fig2}
\end{figure}

{{\it Staggered VBS order in the large $N$ limit.-}} 
In the large-$N$ limit for \Eq{E1}, the mean-field saddle-point approximation is capable of capturing the exact ground-state properties. Hence, before presenting QMC results on finite-$N$ system, we perform self-consistent mean-field calculation to understand the ground-state properties of \Eq{E1} in the large-$N$ limit.  We decouple the interaction in \Eq{E1} in bond channel:  $\varphi_{x/y,i}=\langle (c^{\dagger}_{i}c_{i+\hat x/\hat y}+{\rm{h.c}}) \rangle$ and solve the mean-field equation self-consistently. Our calculation 
indicates sVBS with momentum $(\pi,\pi)$ and cVBS with momentum $(\pi,0)/(0,\pi)$ are two dominant ordered states with close energies. We perform systematic self-consistent mean-field calculation to compare energies of the two states, unequivocally revealing that sVBS ordered state is  energetically favored in the whole interaction parameter regime under consideration. The details and results of mean-field calculation are included in the section V of SM. To conclude, in the large-$N$ limit, the ground state of \Eq{E1} is sVBS ordered phase with momentum $(\pi,\pi)$.  

{{\it AFM order for small $N$.-}} At small or intermediate value of $N$, where most physical realizations reside, quantum fluctuation beyond saddle-point approximation becomes pronounced. Consequently, an approximation-free approach incorporating the effect of quantum fluctuation is strongly desired to investigate the possible SSB at small or intermediate value of $N$. We employ numerically exact QMC simulation to explore the accurate ground-state properties of \Eq{E1}. To demystify various possible symmetry-breaking orders, we compute the static structure factor $S(\vec{q},L)=\frac{1}{L^{4}}\sum_{i,j}e^{i\vec{q}\cdot(\vec{r}_{i}-\vec{r}_{j})}\langle \hat O_{i} \hat O_{j}\rangle$ of the corresponding order $\hat O$ and $L$ refers to the linear system size of square lattice. The order parameter characterizing SSB is given by the peaked-momentum structure factor at thermodynamic limit: $\Delta^{2}=\lim_{L \rightarrow \infty} S(\vec{Q}_{\rm peak},L)$. In our study, we focus on VBS and spin orderings, the definition of which are given in the SM I.  Firstly, we present the momentum distribution of spin structure factor in \Fig{fig2}(a), which clearly displays a sharp peak at momentum $(\pi,\pi)$, implying the existence of AFM ordering at $N=3$. The scaling analysis of AFM structure factor with $1/L$ unambiguously reveals the existence of AFM long-range order, as shown in \Fig{fig2}(b), while the VBS structure factor vanishes at thermodynamic limit. To further confirm the AFM long-range ordered ground state at $N=3$, we perform a more sophisticated finite-size scaling procedure by calculating RG-invariant correlation-length ratio,  $R(L)_{{\rm{AFM/VBS}}}=\frac{S_{{\rm{AFM/VBS}}}(\vec{Q},L)}{S_{{\rm{AFM/VBS}}}(\vec{Q}-\delta\vec{q},L)}-1$, where $\vec{Q}$ labels the momentum at which the structure factor is maximum, $\delta \vec{q}=(\frac{2\pi}{L},\frac{2\pi}{L})$ is a minimal momentum shift from $\vec{Q}$. As shown in \Fig{fig2}(c), the correlation-length ratio of AFM order increases with system size, corroborating the existence of AFM long-range order. Conversely, the system-size dependence of VBS correlation-length ratio, as shown in SM III, is a strong manifestation of short-range property of VBS ordering at $N=3$.  

In addition to the static structure factor, we investigate the spectral properties of \Eq{E1} by virtue of imaginary-time Green's function. At $N=3$, we extract spectral gap from imaginary-time Green's function, with the details illustrated in SM II. The results of single-particle gaps at $N=3$ with varying $J$ are shown in SM III, explicitly demonstrating the existence of finite single-particle gap at thermodynamic limit. Furthermore,  in the AFM ordered phase, the spin gap is expected to vanish at momentum $(\pi,\pi)$ as a consequence of Goldstone modes. \Fig{fig2}(d) depicts the scaling analysis of spin gap as a function of $1/L$, explicitly indicating the absence of spin gap, and hence corroborating the existence of AFM long-range order in the ground state at $N=3$.

\begin{figure}[tb]
\includegraphics[width=0.5\textwidth]{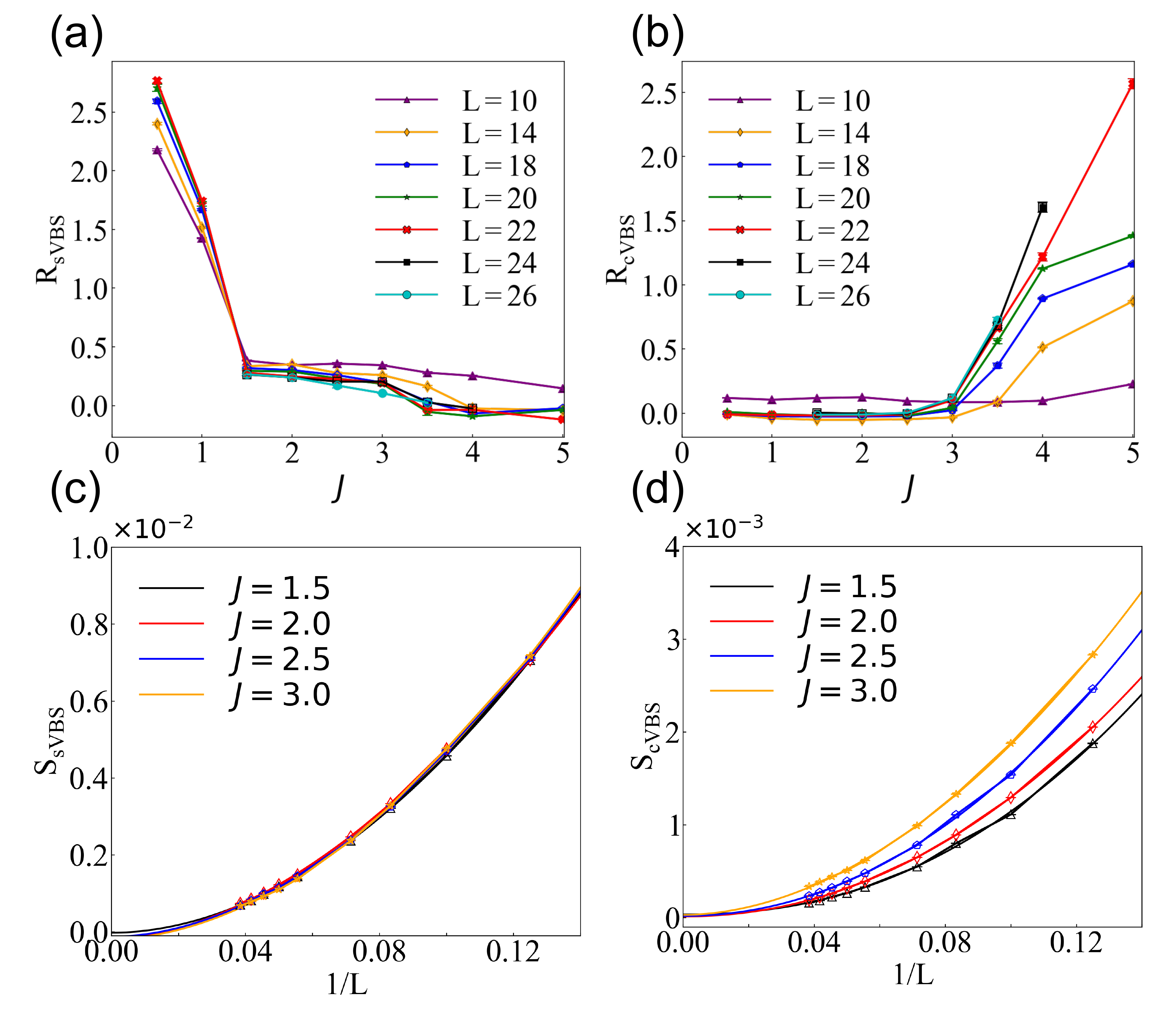}
\caption{The results of correlation-length ratio and structure factors for $N=4$: (a) RG-invariant correlation length ratio of sVBS order. The results increase with system size in the weakly interacting regime $J<1.5$, indicating sVBS ordered phase. (b) Correlation-length ratio of cVBS order. The results increase with system size in the strongly interacting regime $J>3.5$, indicating a cVBS ordered phase. (c) Finite-size scaling of sVBS structure factor in the regime of intermediate interaction strength $1.5 \leq J\leq 3.0 $. (d) Finite-size scaling of sVBS structure factor in the regime of intermediate interaction strength $1.5 \leq J\leq 3.0 $. Both sVBS and cVBS structure factors vanish in thermodynamic limit.  }
\label{fig3}
\end{figure}

{{\it Intermediate Mott insulator without SSB.-}}
In this section, we perform projector QMC to investigate ground-state properties of the model \Eq{E1} in the intermediate values of $N$, wherein the competition between various ordering tendencies is expected to result in exotic phases. We focus our study on $N=4$. For $N=4$, the results clearly show that AFM order is destroyed by the quantum fluctuation of spin by evaluating structure factor and RG-invariant correlation ratio, as shown in section IV of SM. The dominant instability for $N=4$ is VBS ordering. We present the results of correlation-length ratio for sVBS and cVBS orders in \Fig{fig3}(a) and (b), respectively. At weak coupling regime,  the dominant VBS ordering is sVBS with momentum $(\pi,\pi)$ owing to the Fermi surface nesting on square lattice at half filling. Remarkably, in the presence of strong interaction, the ground state displays cVBS order with momentum $(\pi,0)/(0,\pi)$, distinct from the VBS pattern at weak coupling regime. Different from sVBS at weak coupling regime arising from Fermi surface instablity, the cVBS order is a consequence of strong electronic interaction. The results of correlation-length ratio unambiguously establish the existence of sVBS and cVBS long-range order in the region $J<1.5$ and $J>3.5$, respectively. Hence, for the intermediate fermion-component number $N=4$, our simulation unequivocally reveals two different VBS ordered phase emerging in the weak and strong coupling regimes, with distinct patterns and underlying mechanisms.

\begin{figure}[tb]
\includegraphics[width=0.42\textwidth]{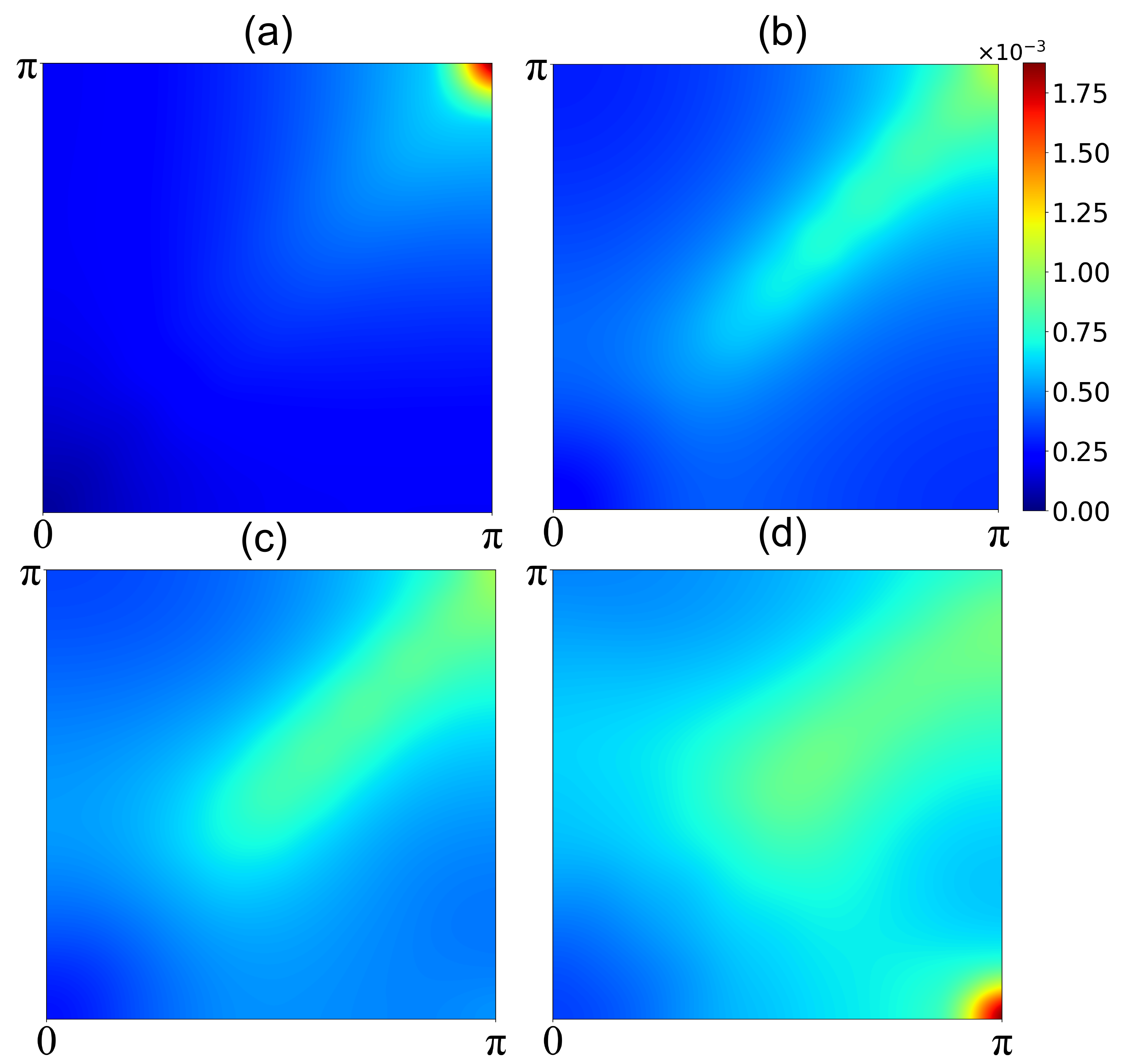}
\caption{Momentum distribution of VBS structure factors for $N=4$: (a) $J=0.5$. (b) $J=2.0$. (c) $J=3.0$.  (d) $J=5.0$. The linear system size is fixed as $L=20$.  }
\label{fig4}
\end{figure}

\begin{figure*}[tb]
\includegraphics[width=0.9\textwidth]{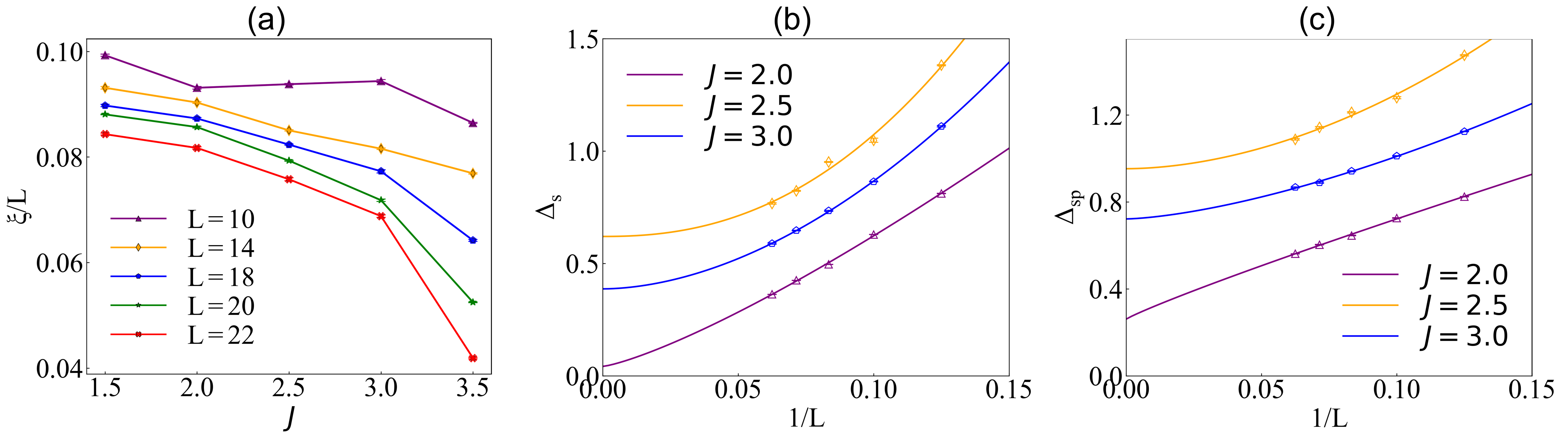}
\caption{(a) Correlation length normalized by linear system size $L$ extracted from RG-invariant correlation ratio for $N=4$. In the intermediate symmetric Mott insulating phase, the correlation length of VBS order is much smaller than $L$. (b) Finite-size scaling of spin gap in the intermediate symmetric Mott insulating phase for $J=2.0,2.5,3.0$ for $N=4$. (c) Finite-size scaling of single-particle gap in the intermediate symmetric Mott insulating phase for $J=2.0,2.5,3.0$ for $N=4$.  } 
\label{fig5}
\end{figure*}

The competition between various ordered phases with incompatible broken symmetries offers a promising route towards yielding exotic QSL phase. In \Eq{E1} for $N=4$, the results of correlation-length ratio reveal an intermediate phase in the region $1.5< J<3.5$, in which both sVBS and cVBS orders are short-range. The finite-size scaling of VBS structure factors with momentum $(\pi,\pi)$ and $(\pi,0)$, as depicted in \Fig{fig3}(c) and (d), also confirm the absence of VBS long-range ordering. We exclude the possibility of other SSB orderings occurring in this intermediate phase by means of finite-scaling analysis of structure factor, as shown in the section IV of SM, corroborating the existence of an intermediate phase without any SSB. The occurrence of intermediate phase in the absence of VBS long-range order is also witnessed by the momentum distribution of VBS structure factors, as shown in \Fig{fig4}. In the weak coupling ($J=0.5$) and strong coupling ($J=5.0$) regimes, as expected, the VBS structure factors are sharply peaked at momentum $(\pi,\pi)$ and $(\pi,0)$, respectively. In the intermediate phase ($J=2.0,3.0$), the peaks of VBS structure factors are obviously broadened, implying the suppression of long-range ordering. To further scrutinize the nature of intermediate symmetric phase, we present the results of correlation length normalized by the linear-system size $L$,estimated from the correlation-length ratio $\xi = \frac{1}{\delta q}\sqrt{\frac{S(Q)}{S(Q+\delta q)}-1}$, in \Fig{fig5}(a). In the intermediate regime $1.5 \textless J \textless 3.5$, the correlation length is much smaller compared with linear system size $L$, confirming that VBS order is short-range and our computational
system size is sufficiently large to capture the physical properties at thermodynamic limit. Taken together, our simulation provides compelling evidences supporting the emergence of an intermediate phase without SSB for $N=4$ in model \Eq{E1}.

To further investigate the nature of intermediate symmetric phase, we calculate the single-particle and spin gaps. The finite-size scaling results of spin and single-particle gaps for various values of $J$ are shown in \Fig{fig5}(b) and (c), respectively. The extrapolated values unambiguously reveal finite single-particle and spin gaps, demonstrating the nature of a gapped Mott insulating phase, and possibly, a gapped quantum spin liquid featuring topological order. Moreover, the intermediate gapped quantum spin liquid phase is also consistent with the Lieb-Schultz-Mattis theorem~\cite{LIEB1961407,oshikawa2000prl,hastings2004prb,yuan2020prx}. In our simulation, the average filling number is two electrons per unit cell, hence the possibility of gapped SU(4) symmetric phase without ground-state degeneracy is excluded owing to the LSM theorem for SU($N$) symmetric system\cite{yuan2020prx}. In conclusion, our systematic simulations establish a possible gapped quantum spin liquid phase existing in the intermediate interacting regime for $N=4$.

{{\it Discussions and concluding remarks:}}  
The SU($N$) interacting fermionic model offers a promising platform to explore the intriguing physics arising from the interplay between strong correlation and enlarged symmetry of multi-component fermions. Here, we perform approximation-free numerical simulation to decipher the ground-state properties of SU($N$) fermionic model with SSH interaction on square lattice. Various competing orderings emerge at varying interaction strength and the number of fermion-component $N$. At small value of $N$, namely $N=2,3$, the ground state is AFM spin ordered state. However, in the large-$N$ limit, the dominant order is sVBS carrying momentum $(\pi,\pi)$ while preserving spin rotational symmetry. The results at intermediate value of $N$ are particularly intriguing. At $N=4$, in weak coupling regime the dominant VBS ordering is sVBS with momentum $(\pi,\pi)$, persisting up to $J=1.5$, whereas the strong interaction between electrons destroy sVBS ordering and establish the cVBS ordered state carrying momentum $(\pi,0)/(0,\pi)$ in region $J>3.5$. More remarkably, in a large intermediate interacting regime ($1.5<J<3.5$), an exotic Mott insulating phase without SSB develops before the cVBS order is established. The intermediate symmetric Mott insulating phase possesses finite single-particle and spin gaps. All these features are consistent with a gapped quantum spin liquid phase. 

Notably, although we focus on the simulation on model in \Eq{E1} at half filling, the model is sign-problem-free at generic filling for even $N$. The possible superconducting pairing arising from doping various symmetry spontaneously breaking phases or quantum spin liquid is of particular interest. Furthermore, \Eq{E1} is sign-problem-free at half filling for even $N$ in the presence of repulsive Hubbard interaction, making it feasible to investigate the effect of Hubbard interactions on the various VBS and QSL order unveiled in our simulation. Another interesting direction to explore in future is the nature of quantum phase transition between quantum spin liquid and different VBS ordered phases. Therefore, our simulation paves a promising way to constructing concrete theoretically tractable model to investigate these crucial ingredients of quantum many-body physics in a unified framework.

{{\it Acknowledgement.-}}  We thanks Hong Yao, Yi Zhou,  and Linhao Li for very helpful discussions. We thank the computational resources provided by the TianHe-1A supercomputer, the High Performance Computing Platform of Peking University, China. The QMC simulations are partly carried out with ALF Library\cite{alf_assaad}. X.-J.Y. and L.X. are supported by the National Natural Science Foundation of China under Grant No.11935002, and the National 973 project under Grant No. 2021YF1400501. Z.X.L acknowledges support from the
start-up grant of IOP-CAS.

\bibliography{SUNfermions}

\setcounter{equation}{0}
\setcounter{figure}{0}
\setcounter{page}{1}
\renewcommand{\theequation}{S\arabic{equation}}
\renewcommand{\thefigure}{S\arabic{figure}}
\def\avg#1{\left\langle#1\right\rangle}

\newpage

\begin{widetext}

\section*{Supplemental Material}

\renewcommand{\theequation}{S\arabic{equation}}
\setcounter{equation}{0}
\renewcommand{\thefigure}{S\arabic{figure}}
\setcounter{figure}{0}
\renewcommand{\thetable}{S\arabic{table}}
\setcounter{table}{0}

\subsection{I. The details of projector Quantum Monte Carlo simulation}
\label{sec:SM1}
We employ the method of projector QMC to investigate the ground-state properties of the SU($N$) SSH model at the anti-adiabatic limit, as described in ~\Eq{E1} in the main text. In Projector QMC, the ground-state expectation value of the observable is evaluated by projecting a trial wave function $\ket{\psi_T}$ along the imaginary time direction:

\begin{equation}
\label{S1}
\langle {\hat{O}} \rangle = \frac{\bra {\psi_0} \hat{O} \ket {\psi_0}}{\langle \psi_0 \mid \psi_0 \rangle} = \lim_{\Theta \rightarrow \infty} \frac{\bra{\psi_T}e^{-\frac{\Theta H}{2} } \hat{O} e^{-\frac{\theta H}{2}}\ket{\psi_T}}{\bra{\psi_T}e^{- \Theta H}\ket{\psi_T}} 
\end{equation}
Here $\ket{\psi_T}$ is the trial wave function and $\Theta$ is the projection parameter. The key point of the algorithm is intrinsically unbiased against the choice of $\ket{\psi_T}$ assuming that the trivial wave function is not orthogonal to the exact ground state, namely $\langle \psi_T \mid \psi_0 \rangle \neq 0$, which is generically satisfied for quantum many-body models in finite systems. In our simulation, we choose $\ket{\psi_T}$ as the ground-state wave function of the non-interacting part of the model under consideration. The projection parameter $\Theta$ in \Eq{S1} is chosen to be $\Theta=40$ for most simulations in our study. For the simulations of model \Eq{E1} for $N=4$, we choose a larger projection parameter $\Theta=60$ to determine the ground-state properties of competing ordered phases and the possible quantum spin liquid phase. We have checked that the value of $\Theta$ is sufficient large to access the ground-state properties of the model.  

Similar to the finite-temperature determinant QMC algorithm, in projector QMC Trotter decomposition is implemented to discretize the imaginary time $\Theta$ in spacing $\Delta_\tau=\Theta/L_{\tau}$, introducing a Trotter error scaling as $\Delta_\tau^2$.  In our simulation, we choose $\Delta_\tau=0.1$ for most cases and $\Delta_\tau = 0.05$ for the simulation of model \Eq{E1} for $N=4$. Of course, We have checked the convergence of the results against further decreasing the value of $\Delta_\tau$. To decouple the four-fermion interacting terms in the SU($N$) model \Eq{E1}, we perform discrete Hubbard-Stratonovich transformation via introducing a classical space-time dependent auxiliary field after the procedure of Trotter decomposition:
\bea
e^{\frac{\Delta_\tau J}{2 N}(\sum_\alpha c^\dagger_{i\alpha} c_{j\alpha} + H.c. ) ^2} = \sum_{l=\pm 1,\pm 2} \gamma(l) e^{\sqrt{\frac{\Delta_\tau J}{2N}}\eta(l)(\sum_\alpha c^\dagger_{i\alpha} c_{j\alpha} + H.c.)} + O(\Delta_\tau^4)
\eea
with the four-valued parameters introduced as: $\gamma(\pm 1) = 1+\sqrt{6}/3$, $\gamma(\pm 2) = 1-\sqrt{6}/3$, $\eta(\pm 1) = \pm \sqrt{2(3-\sqrt{6})}$, $\eta(\pm 2) = \pm \sqrt{2(3+\sqrt{6})}$. This scheme of Hubbard-Stratonovich transformation keeps the SU($N$) symmetry, and is utilized to decouple the SU($N$) four-fermion interactions.

To characterize various long-range orderings, we evaluate structure factors for the corresponding order parameters. For model \Eq{E1}, the dominant orderings are AFM spin order breaking spin rotational symmetry and VBS order breaking lattice transnational symmetry. The structure factors for spin and VBS orders with system size $N_s=L\times L$ are defined as: 
\begin{equation}
\label{S2}
S_{\rm{spin}}(L,\vec{k})=\frac{1}{N_s^2N^2}\sum_{i,j,\alpha,\beta} \langle \hat{S}_{\alpha\beta}(i)\hat{S}_{\beta\alpha}(j) \rangle e^{i\vec{k}\cdot (\vec{r_{i}}-\vec{r_{j}})} 
\end{equation}
\begin{equation}
\label{S32}
\begin{split}
&S_{{\rm{VBS}}}(L,\vec{k},\delta) = \frac{1}{N_s^2 N^2}\sum_{i,j} \langle \hat{K}_{\delta}(i)\hat{K}_{\delta}(j)\rangle e^{i\vec{k}\cdot (\vec{r_{i}}-\vec{r_{j}})},
\end{split}
\end{equation}
where $\vec{k}$ is the momentum of structure factor in consideration. For spin order, $\hat{S}_{\alpha\beta}(i) = c^\dagger_{i\alpha}c_{i\beta}-\frac{\delta_{\alpha\beta}}{N}\sum^N_{\gamma=1} c^\dagger_{i\gamma}c_{i\gamma} $($\alpha,\beta=1,\cdots,N$) is SU($N$) spin operator on site $i$, representing the generators of SU($N$) group. Because of the relation $\sum_\gamma S_{\gamma\gamma}=0$, there are $N^2-1$ independent generators for the SU($N$) group. AFM order on square lattice denotes the spin density wave ordering with momentum $\vec{k}=(\pi,\pi)$, hence the definition of structure factor for AFM order reads: $S_{\rm{AFM}}(L)= S_{\rm{spin}}(L,\vec{k}=(\pi,\pi))$. For VBS structure factor, $\hat{K}_{\delta}(i)=\sum_\alpha c^{\dagger}_{i\alpha}c_{i+\delta\alpha}+H.c.$ is the kinetic operator on bond in $\delta=x,y$ direction. For \Eq{E1}, two distinct VBS orderings are dominant with varying interaction strength $J$ and fermion flavor number $N$, namely sVBS and cVBS. The sVBS order carries momentum $\vec{k}=(\pi,\pi)$. In our simulation, we evaluate structure factor for sVBS order averaging over $\hat{x}$ and $\hat{y}$-direction, which is defined as $S_{\rm{sVBS}} = \frac{1}{2}(S_{\rm{VBS}}(L,\vec{k}=(\pi,\pi),\hat{x})+S_{\rm{VBS}}(L,\vec{k}=(\pi,\pi),\hat{y}))$. The cVBS order carries momentum $\vec{k}=(\pi,0)$ for kinetic term in $\hat{x}$-direction and momentum $\vec{k}=(0,\pi)$ for kinetic term in $\hat{y}$-direction. The structure factor for the cVBS order is defined as $S_{\rm{cVBS}}= \frac{1}{2}(S_{\rm{VBS}}(L,\vec{k}=(\pi,0),\hat{x})+S_{\rm{VBS}}(L,\vec{k}=(0,\pi),\hat{y}))$. VBS order breaks the lattice $\mathbb{Z}_4$ symmetry, while AFM breaks spin $SU(2)$ rotational symmetry. The order parameter characterizing SSB is given by the corresponding structure factor at thermodynamic limit: $\Delta^2 = {\rm{lim}}_{L \rightarrow \infty} S(L)$.

\subsection{II. Derivations of single-particle and spin gap from time-displaced correlation function}
\label{sec:SM2}
To understand the spectral properties of the model, we evaluate the single-particle excitation gap extracted from the imaginary-time displaced Green's function. The single-particle gap denotes the minimum energy cost to extract one electron from the system, corresponding to the spectral gap detected in the photoemission experiment. Moreover, we extract the spin gap $\Delta_{\rm{s}}(\textbf{k})$ from imaginary-time displaced spin-spin correlation function. The spin gap is expected to vanish in the AFM ordered phase owing to the existence of Goldstone mode. 

To extract single-particle gap, we compute SU($N$) imaginary-time displaced single-particle Green's function:
\bea
G(\vec{k},\tau) = \sum_\alpha \avg{c_{\vec{k}\alpha}(\tau) c^\dagger_{\vec{k}\alpha}(0)} = \frac{1}{N^2_s}\sum_\alpha  \sum_{ij}\avg{c_{i\alpha}(\tau)c^\dagger_{j\alpha}(0)} e^{i(\vec{R}_i-\vec{R}_j)\cdot \vec{k}}
\eea
where $\vec{k}$ is the momentum under consideration and $N_s$ is the total number of lattice site. The asymptotic behaviour at large $\tau$ gives rise to the estimation of single-particle gap at momentum $\vec{k}$: $G(\vec{k},\tau) \sim e^{-\Delta_{\rm{sp}}(\vec{k})\tau}$. Similarly, spin gap is extracted from the imaginary-time displaced correlation function of spin operator:
\bea
G_s(\vec{k},\tau) = \sum_{\alpha \beta} \avg{S_{\alpha \beta}(\vec{k},\tau) S_{\beta \alpha}(\vec{k},0)}
\eea
where $S_{\alpha \beta}(i)$ is SU($N$) spin operator defined as $S_{\alpha \beta}(i) = c^\dagger_{i\alpha}c_{i\beta}-\frac{\delta_{\alpha\beta}}{N}\sum^N_{\gamma=1} c^\dagger_{i\gamma}c_{i\gamma}$ and $S_{\alpha \beta}(i\tau) = e^{\tau H} S_{\alpha \beta}(i) e^{-\tau H}$. $\vec{k}$ is the momentum under consideration and $N_s$ is the total number of lattice sites. At sufficient large $\tau$, the spin correlation function scales as $G_s(\vec{k},\tau) \sim e^{-\Delta_{\rm {spin}}(\vec{k}) \tau}$. Hence, the single-particle gap and spin gap at a given momentum are available to extract from the single-particle Green's function and spin correlation function at large imaginary time $\tau$ respectively. In our simulation, we present the results of single-particle gap and spin gap with the minimum value of $\Delta_{\rm{sp}}(\vec{k})$ and $\Delta_{\rm{spin}}(\vec{k})$ in the whole Brillouin zone. For the results of spin gaps for $N=3,4$ as shown in main text, the minimum spin gap $\Delta_{\rm{spin}}(\vec{k})$ occurs at $\vec{k}=(\pi,\pi)$ in the whole interaction regime under consideration. For single-particle gap, we compute local real-space imaginary-time displaced Green's function equivalent to the average of $G(\vec{k},\tau)$ in the whole Brillouin zone: $G(i,\tau) = \frac{1}{N_s}\sum_{i,\alpha} \avg{c_{i\alpha}(\tau)c^\dagger_{i\alpha}(0)} = \frac{1}{N_s} \sum_{\vec{k}} G(\vec{k},\tau) $. At sufficient long imaginary time, $G(i,\tau)$ is determined by the minimum value of $\Delta_{\rm{sp}}(\vec{k})$ in the whole Brillouin zone, scaling as $G(i,\tau) \sim e^{-\Delta_{\rm{sp}}\tau}$  where $\Delta_{\rm{sp}}$ is the spin gap defined as the minimum gap in the whole Brillouin zone. Hence, we extract single-particle gap from the scaling behaviour of $G(i,\tau)$ at long imaginary time.

\subsection{III. Additional QMC results for $N=3$}
\label{sec:SM3}

In this section, we present additional QMC results of \Eq{E1} for $N=3$, substantiating the conclusion that the ground state is AFM long-ranged ordered phase for $N=3$.

In the maintext, we provide convincing evidence that the ground state possesses AFM long-range order. Here we present the results of VBS ordering, confirming the absence of VBS long-range order in the ground state for $N=3$. For $N=3$, the peaked momentum of VBS structure factor is $(\pi,\pi)$, hence we evaluate correlation-length ratio of staggered VBS order versus $J$ for various system sizes, as shown in \Fig{figs1}(a). Additionally, we compute the single-particle gap at several values of $J$ for $L=8,10,12,14,16$, and fit the results using second-polynomial function of $1/L$, the intercepts of which give rise to the single-particle gap at thermodynamic limit. The results of single-particle gap are depicted in \Fig{figs1}(b), clearly demonstrating the single-particle gaps are opened in the ground state for $N=3$.

\subsection{IV. Additional QMC results for $N=4$}
\label{sec:SM4}
In the maintext, we present QMC numerical results for VBS order and establish the ground-state phase diagram for $N=4$. In the weakly coupling regime, the ground state is a sVBS ordered phase, whereas in the strongly coupling regime the dominant VBS ordering is columnar order carrying momentum $(\pi,0)/(0,\pi)$. More remarkably, there exists an intermediate disordered phase preserving all the lattice and Hamiltonian symmetries between the staggered and columnar VBS ordered phases in the weakly and strongly coupling regime. In this section, we present additional data of QMC simulation for $N=4$, to exclude the possibility for the existence of long-range ordered phases. The results of AFM correlation-length ratio and structure factors are shown in \Fig{figs2}(a) and (b), respectively. The correlation-length ratio of AFM order decreases with linear system size $L$ in the whole interacting parameter regime under consideration, indicating the AFM order is short-ranged for $N=4$. In addition, the fitting results of AFM structure factors further confirm the absence of AFM long-range order in the ground state for $N=4$. 

For even value of $N$, the model \Eq{E1} on bipartite lattice  features an additional partial particle-hole $Z_2$ symmetry exists at half filling: $c_{i\alpha} \rightarrow c^\dagger_{i\alpha} (-1)^{i_x+i_y}$ for even $\alpha$. Under such partial particle-hole transformation, the diagonal spin operator $\sum_\alpha c^\dagger_{i\alpha}c_{i\alpha} (-1)^\alpha$ is transformed to the density operator $n_i = \sum_\alpha c^\dagger_{i\alpha}c_{i\alpha}$. The off-diagonal spin operator $c^\dagger_{i\alpha}c_{i\alpha+1} $ ($\alpha$ is odd) is transformed to staggered pairing operator $c^\dagger_{i\alpha}c^\dagger_{i\alpha+1}(-1)^{(i_x+i_y)}$. Consequently, the AFM order parameters are degenerate with CDW order parameter $O_{\rm{CDW}} = \frac{1}{N_s} \sum_i n_i (-1)^{i_x+i_y}$ and on-site superconducting order parameter $O_{\rm{SC}}= \frac{1}{N_s} \sum_i c^\dagger_{i\alpha}c^\dagger_{i\alpha+1} $. Because we have unambiguously shown the absence of AFM order in the ground state for $N=4$, the CDW and on-site singlet SC pairing order are also short-ranged in the ground state of model \Eq{E1} for $N=4$.

\begin{figure}[tb]
\includegraphics[width=0.8\textwidth]{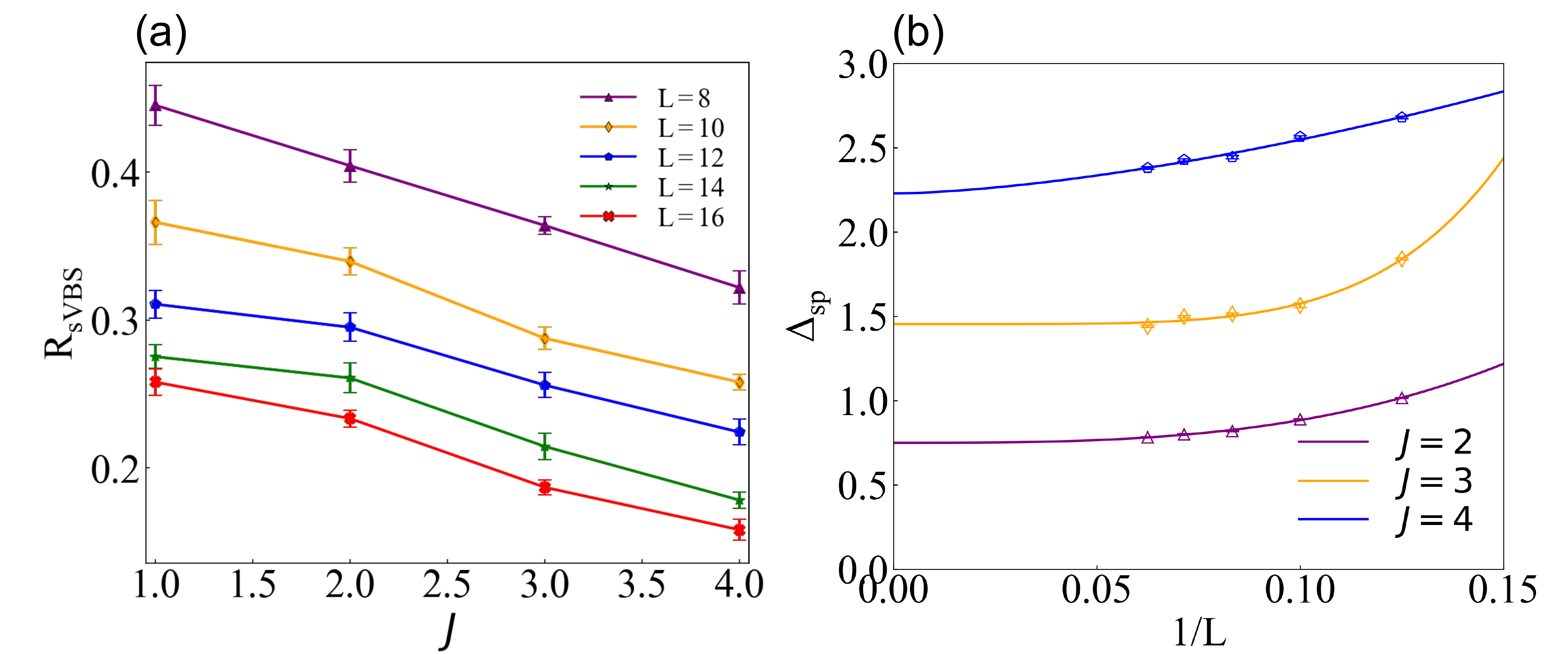}
\caption{(a) Correlation-length ratio of sVBS order parameters $R_{\rm{sVBS}}$ as a function of interacting strength $J$ for $N=3$. Correlation-length ratio $R_{\rm{sVBS}}$ decreases with linear system size $L$, indicating the staggered VBS long-range order is absent. (b)The results of single-particle gap as a function of $1/L$ for $J=2,3,4$ and $N=3$. The extrapolation of single-particle gap as a function of $1/L$ to thermodynamic limit is performed, which unambiguously reveal the finite single-particle gap exists in the AFM ordered phase for $N=3$. }
\label{figs1}
\end{figure}

\begin{figure}[tb]
\includegraphics[width=0.8\textwidth]{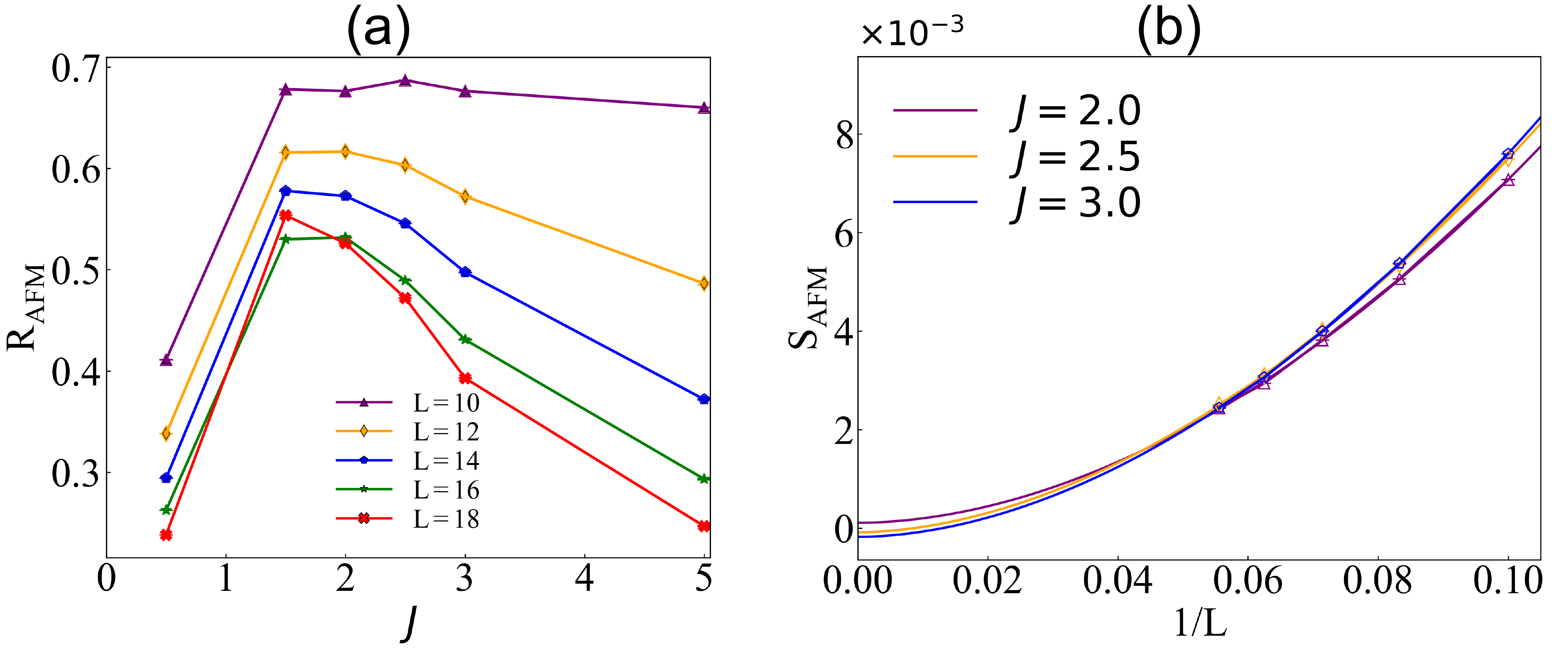}
\caption{(a) Correlation-length ratio of AFM order parameters $R_{\rm{AFM}}$ as a function of interacting strength $J$ for $N=4$. Correlation-length ratio $R_{\rm{AFM}}$ decreases with linear system size $L$, indicating the absence of AFM long-range order. (b) The finite-size scaling analysis for AFM structure factor as a function of $1/L$ for $J=2.0,2.5,3.0$ and $N=4$. The extrapolated value at thermodynamic limit $1/L \rightarrow \infty$ within error bar, further confirming the absence of AFM long-range order in the ground state for $N=4$.}
\label{figs2}
\end{figure}

\subsection{V. Staggered VBS ordered phase in the large $N$ limit}
\label{sec:SM5}
For model \Eq{E1}, in the large-$N$ limit, the saddle-point approximation, namely mean-field approximation, is able to give rise to the exact solution since the fluctuation effect beyond saddle-point approximation is suppressed by $\frac{1}{N}$. Here we perform self-consistent mean-field calculation to investigate the ground-state properties of the model. The interaction term reads:
\bea
H_J = -\frac{J}{2N}\sum_{\langle ij\rangle}(\sum_\alpha c^{\dagger}_{i\alpha}c_{j\alpha}+h.c.)^2
\eea
We consider SU($N$) symmetric mean-field decoupling of the interaction in terms of bond hopping mean-field value: 
\bea
\varphi(i,\hat{\delta}) = \langle  c^{\dagger}_{i\alpha} c_{i+\hat{\delta}\alpha} + h.c.\rangle, 
\label{meanfield}
\eea
resulting in the mean-field Hamiltonian:
\begin{equation}
\label{S9}
H^{\rm{MF}}=-\sum_{i\alpha}(t+J\varphi(i,\hat{x}))(c^{\dagger}_{i\alpha} c_{i+\hat{x}\alpha}+h.c.)-\sum_{i\alpha}(t+J\varphi(i,\hat{y}))(c^{\dagger}_{i\alpha} c_{i+\hat{y}\alpha}+h.c)+\sum_{i}\frac{J}{2}(\varphi^2(i,\hat{x})+\varphi^2(i,\hat{y}))
\end{equation}
Since the mean-field Hamiltonian $H^{\rm{MF}}$ is quadratic and can be diagonalized straightforwardly. Thus, we can solve the mean-field equation \Eq{meanfield} self-consistently. Notice that given the $SU(N)$ symmetric ansatz \Eq{meanfield}, the results of mean-field solution is independent on the value of $N$. We consider two mean-field ansatz corresponding to the staggered and columnar VBS orders. For the sVBS order:
\bea
\langle  c^{\dagger}_{i\alpha} c_{i+\hat{x} \alpha} + h.c.\rangle &=& \phi^s_0 + \phi_{\rm{stag}} (-1)^{x+y} \nonumber\\
\langle  c^{\dagger}_{i\alpha} c_{i+\hat{y}\alpha} + h.c.\rangle &=& \phi^s_1
\eea
For the cVBS order:
\bea
\langle  c^{\dagger}_{i\alpha} c_{i+\hat{x} \alpha} + h.c.\rangle &=& \phi^c_0 + \phi_{\rm{colu}} (-1)^{x} \nonumber\\
\langle  c^{\dagger}_{i\alpha} c_{i+\hat{y} \alpha} + h.c.\rangle &=& \phi^c_1
\eea
where $\phi_{\rm{stag}}$ and $\phi_{\rm{colu}}$ are order parameters characterizing staggered and columnar VBS orders respectively. $\phi^s_0$, $\phi^s_1$, $\phi^c_0$ and $\phi^c_1$ represent uniform parts in the mean-field ansatz. The values of these parameters are solved self-consistently. The results of sVBS and cVBS order parameters as a function of interaction strength $J$ are shown in \Fig{figs3}(a). The sVBS order parameter is finite in the whole interacting parameter regime under consideration, increasing with interacting strength $J$. On the other hand, for cVBS order, the self-consistent mean-field calculations show that order parameter remains zero in a large interacting parameter regime $J \leq 4$. Given by the mean-field solutions, the ground-state energy of the staggered and columnar VBS phases for mean-field Hamiltonian are easily achieved. We present the results of ground-state energies for staggered and columnar VBS phases in \Fig{figs3}(b), unambiguously indicating staggered VBS phase is energetically favored in the whole interaction parameter regime under consideration. Hence, the mean-field calculations provide convincing evidence that the ground state of model \Eq{E1} is sVBS ordered phase in the large-$N$ limit.

\begin{figure}[tb]
\includegraphics[width=0.8\textwidth]{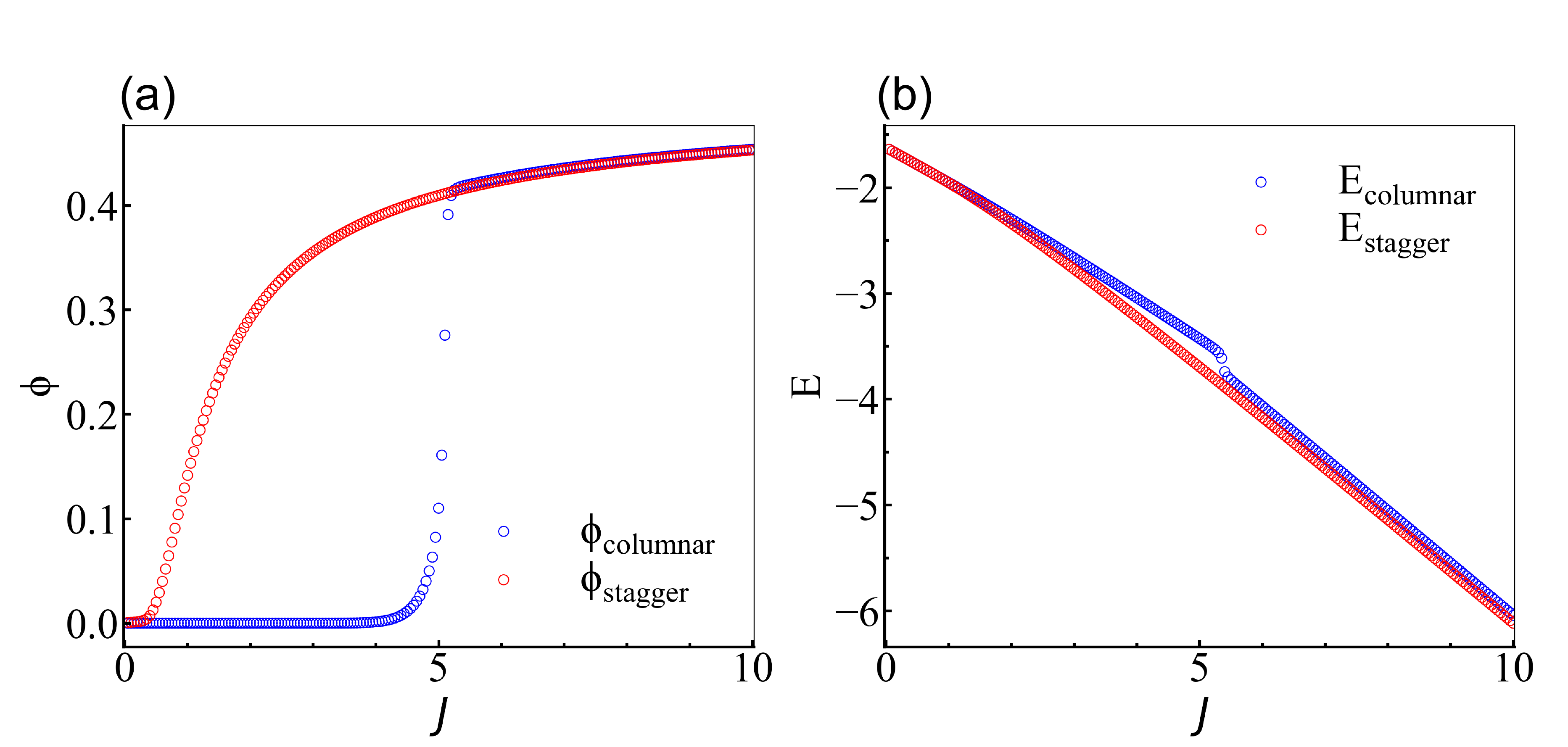} 
\caption{Large $N$ mean field results for SU(N) SSH model. (a) Staggered and columnar VBS order parameters as a function of interaction strength $J \in [0,10]$. (b) Results of ground-state energies for staggered and columnar VBS phase as a function of interaction strength $J \in [0,10]$.}
\label{figs3}
\end{figure}

\end{widetext}

\end{document}